\documentclass[prb]{revtex4}
\usepackage{dcolumn,amsmath}
\usepackage[dvips]{epsfig}
\begin{document}
\title{Multicanonical Parallel Tempering}
\author{Roland Faller}
\author{Qiliang Yan}
\author{Juan J. \surname{de Pablo}}
\affiliation{Department of Chemical Engineering, University of Wisconsin,
Madison, WI 53706}
\renewcommand{\thetable}{\Roman{table}}
\newcommand{\mc}[1]{\multicolumn{1}{c}{#1}}
\begin{abstract}
We present a novel implementation of the parallel tempering Monte Carlo method
in a multicanonical ensemble. Multicanonical weights are derived by a 
self-consistent iterative process using a Boltzmann inversion of global energy 
histograms. This procedure gives rise to a much broader overlap of 
thermodynamic-property histograms; fewer replicas are necessary in parallel 
tempering simulations, and the acceptance of trial swap moves can be made
arbitrarily high. We demonstrate the usefulness of the method in the context of
a grand-multicanonical ensemble, where we use multicanonical simulations in 
energy space with the addition of an unmodified chemical potential term in 
particle-number space. Several possible implementations are discussed, and the 
best choice is presented in the context of the liquid-gas phase transition of 
the Lennard-Jones fluid. A substantial decrease in the necessary number of 
replicas can be achieved through the proposed method, thereby providing a 
higher efficiency and the possibility of parallelization.
\end{abstract}
\maketitle
\section{Introduction}

Advanced Monte-Carlo simulation techniques can facilitate considerably the
study of complex systems. Two classes of methods that have proven to be
particularly useful are parallel tempering
techniques~\cite{marinari92,tesi96,hansmann97,wu99,yan99} (sometimes also
called multiple Markov chains), and multicanonical
techniques~\cite{torrie77,berg92a,berg00}. A recent review of these methods
can be found in the literature~\cite{iba01}.

In parallel tempering, several independent replicas of a system are simulated
simultaneously. Each replica (or simulation box) can experience different
thermodynamic-state conditions (e.g. temperature, pressure or chemical
potential). Neighboring systems (in the sense that their state points are not
too distant from each other) are allowed to interchange configurations from
time to time, subject to specific acceptance criteria. These so-called ``swap''
moves can improve sampling of configuration space considerably, particularly in
systems having rugged energy landscapes. Replicas of a system which are close
to a glassy state, for example, may exchange their way ``up'' in, say,
temperature, to states where energy barriers are easier to overcome; they can
subsequently come back to low temperatures to yield an uncorrelated
configuration.

More specifically, $t$ {\it independent} replicas of the same system are 
simulated under different thermodynamic conditions, $C_1, \;C_2,\ldots, \;C_t$,
where the $C_i$ denote combinations of intensive variables (e.g. temperature 
and chemical potential) which differ from replica to replica. Conventional 
Monte Carlo trial moves are conducted in each replica $i$ to sample 
configuration space. In addition, trial swap moves involving two replicas $i$ 
and $j$ are also attempted; in these trial moves, entire conformations are 
interchanged. The acceptance criteria for a trial swap can be derived from the 
product of the elementary moves which are used to construct it. We use 
subscripts $m$ and $n$ to denote the configurations pertaining to two distinct
replicas, or simulation boxes; $i$ and $j$ are used to denote their respective
thermodynamic conditions. In the particular case of a grand-canonical ensemble,
the probability of accepting an individual trial move from configuration $m$ to
$n$ at reduced inverse temperature $\beta_i=1/k_BT_i$ and chemical potential
$\mu_i$ is given by:
\begin{equation}
  p_i=\min\{1,\exp[-\beta_i(E_{n}-E_{m})+\beta_i\mu_i(N_{n}-N_{m})]\},
\end{equation}
where $N_{n}$ denotes the number of particles in replica $n$. That of accepting
an individual trial move from configuration $n$ to $m$ at temperature $T_j$ and
chemical potential $\mu_j$ is given by:
\begin{equation}
  p_j=\min\{1,\exp[-\beta_j(E_{m}-E_{n})+\beta_j\mu_j(N_{m}-N_{n})]\}.
\end{equation}
A swap move can be viewed as a ``double-move'', for which the acceptance
probability is the product of $p_i$ and $p_j$:
\begin{eqnarray}
p_{ij}&=&\min\{1,\exp[-\beta_i(E_{n}-E_{m})-\beta_j(E_{m}-E_{n})+\nonumber\\
&& \beta_i\mu_i(N_{n}-N_{m})+\beta_j\mu_j(N_{m}-N_{n})]\}
\nonumber\\
&=&\min\{1,\exp[-(\beta_i-\beta_j)(E_{n}-E_{m})+
(\beta_i\mu_i-\beta_j\mu_j)(N_{n}-N_{m})]\}. \label{eq:metropolis}
\end{eqnarray}

From Equation~\ref{eq:metropolis} it can be seen that swap trial moves are 
only accepted if some degree of overlap exists between the probability 
distribution functions (or histograms) corresponding to neighboring state 
points (or replicas). One shortcoming of parallel tempering is that the number 
of replicas required for an effective simulation increases with the size of the
simulated system. This is a result of the central limit theorem, which shows 
that the width of thermodynamic-property histograms scales as $\sqrt{N}^{-1}$,
thereby decreasing the extent of overlap between histograms corresponding to 
neighboring replicas.

Parallel tempering simulations improve sampling by shuttling configurations 
from regions of low temperature or high chemical potential to regions of high 
temperature or low chemical potential, where a system can relax more easily. 
They have the added feature that each of the replicas generates useful
information (e.g. thermodynamic quantities, structure) about the system of 
interest. Multicanonical simulations follow an entirely different strategy to 
overcome high-energy barriers between neighboring, local free-energy minima: 
the acceptance criteria for the transition between two states are manipulated 
in such a way as to artificially lower such barriers. The conventional energy
distribution for a canonical ensemble involves two contributions: the density 
of states $\Omega (E)$ and a Boltzmann, exponential energy weight of the form 
$\exp[-\beta E]$. On the one hand, the density of states increases rapidly with
energy and system size and, on the other hand, the exponential energy term 
leads to a suppression of high energy states when the energy exceeds the
thermal energy significantly. The product of the density of states and the 
Boltzmann weight therefore results in a Gaussian-like energy distribution. In 
multicanonical simulations, the conventional Boltzmann weight is replaced by a 
different, non-Boltzmann weight $w_{\text{NB}}$, which is conceived in such a
way as to result in a flat energy distribution. A flat distribution would be 
desirable for two reasons: from a statistical-mechanics point of view, 
realizing a perfectly flat energy distribution is equivalent to calculating the
density of states of the system (or its microcanonical-ensemble partition 
function); the logarithm of this quantity is the entropy. From a more technical
point of view, realizing a flat energy distribution ensures that all states are
sampled with comparable frequency, thereby improving statistics.

For concreteness, the following discussion is restricted to one-dimensional 
distributions depending only on energy; the extension to multidimensional cases
is straightforward. The probability $p(E)$ of finding the system of interest in
a given energy state can be expressed in the form
\begin{equation}
  p(E) = \Omega (E) \; w(E)\label{eq: general} .
\end{equation}
Equation~(\ref{eq: general}) is valid regardless of the weights $w(E)$; 
different weights characterize different ensembles. For a canonical, 
$NVT$-ensemble, $w_{NVT}(E)=\exp(-\beta E)$. In multicanonical simulations a 
final set of weights is calculated in such a way as to make $p(E)$ flat, 
\emph{i.e.} independent of $E$. A perfectly flat distribution could be 
generated if the following weights were employed
\begin{equation}
  w(E) = \frac{1}{\Omega(E)}=\exp(-S(E)/k_B) \label{eq:muka},
\end{equation}
where $S(E)$ is the entropy as a function of energy. In view of the form of
Eqn.(\ref{eq:muka}), simulation techniques in which different energy states are
sampled with uniform probability are sometimes referred to as entropic sampling
methods.

To calculate properties using a uniform-energy, or multicanonical sampling 
technique, energy histograms $H_{\text{mc}}(E)$ are generated during the course
of the simulation. These histograms provide estimates of the probability of 
finding a configuration having energy $E$ in a multicanonical ensemble; they 
can subsequently be ``reweighted'' in order to generate results in one of the 
more conventional ensembles. The so-called multicanonical weights, which 
dictate the sampling, are denoted by $w_{\text{mc}}(E)$. Their construction is 
discussed later in this work. Taking the internal energy as an example, we have
\begin{equation}
  \langle E \rangle_{\text{can}} = \sum_E
  \frac{\exp(-\beta E)}{w_{\text{mc}}(E)}H_{\text{mc}}(E) E,
\end{equation}
where the brackets denote a canonical ensemble average.

The two methods, parallel tempering and multicanonical simulations, have 
several attributes of their own. It is therefore of interest to explore the 
possible advantages of a combined method, in which parallel tempering would be
used to have independent random walkers in different parts of the energy
landscape, and multicanonical Monte Carlo would be employed to reduce the 
number of necessary replicas. This work investigates such a combination. Ideas
similar in spirit were pursued by Sugita {\it et al.}~\cite{sugita00a} and 
Calvo and Doye~\cite{calvo00}. Sugita's work, however, only considered 
single-molecule simulations. It is unclear whether the Sugita {\it et al.}
approach would be of use in a many-body system. The work of Calvo and Doye 
proposes a scheme that differs from ours in that it involves exchanges between 
one multicanonical trajectory and multiple tempering replicas. Furthermore, 
that work is also limited to single molecules or small atomic clusters, and
therefore does not address many of the issues that arise in many-body, 
condensed phases.
\section{Multicanonical parallel Tempering}
In this work, we have chosen to implement a multicanonical sampling scheme 
through a so-called {\it umbrella} potential $\xi(E)$, which is added to the 
energy in the grand canonical-ensemble. Our multicanonical weights are of the 
form
\begin{equation}
  w_{\text{mc}}(E,N) = \exp[-\beta(E+\xi(E))+\beta\mu N] .
\end{equation}
The factor $\exp(-\beta\xi(E))$ changes the distribution from Boltzmann to one
possible type of multicanonical. In order to satisfy Equation~(\ref{eq:muka}),
the ideal umbrella potential $\xi(E)$ should be of the form
\begin{equation}
  \xi(E) = -E +TS(E). \label{eq:free}
\end{equation}
This would lead to purely entropic sampling in that all energy states would be
visited with equal frequency, according to $1/\Omega(E)$.

Unfortunately, the entropy of a system to be simulated is not known \'a priori;
the calculation of $\xi$ must therefore be carried out through a 
self-consistent, iterative process. A series of simulations are conducted; the
umbrella potential is adjusted in such a way as to render the energy landscape
corresponding to each simulation successively flatter, i.e. the ``weights'' of
formerly poorly visited states are augmented, and those of more heavily visited
states are reduced.

Our starting point is a grand-canonical, multi-dimensional parallel tempering 
simulation, where we set $\xi^{(0)}(E)=0$ over the entire energy range. Upper 
indices refer to iteration numbers. We use one single, global $\xi(E)$ for all 
replicas. Otherwise, an uncontrolled bias would lead to incorrect estimates of
the histograms and ultimately incorrect results. After a few thousand 
simulation cycles, we analyze the global energy histogram derived from all the
replicas
\begin{equation}
  H(E) = \sum_i H_i(E)\,,
\end{equation}
where $H_i(E)$ is the energy histogram collected in replica $i$. This 
histogram is now ``Boltzmann inverted'' (Eq.~(\ref{eq:binv})), i.e. its 
logarithm is multiplied by $k_BT$ to obtain an estimate of the corresponding 
weight. The current value of $\xi^{(n)}(E)$ is then updated according to:
\begin{eqnarray}
  \xi^{(n+1)}(E)& = &\xi^{(n)}(E) + \beta_{\text{max}}^{-1}\ln H(E)-
  \beta_{\text{max}}^{-1}\overline{\ln H(E)}.\label{eq:binv}
\end{eqnarray}
The third term in Equation~(\ref{eq:binv}) is a constant, and it corresponds to
the average over all of the $\ln H(E)$; it drops out of any acceptance 
criteria. Its sole purpose is computational efficiency. It allows the umbrella
potentials to increase as well as decrease between iterations (its omission 
leads to more iterations).

\begin{figure}
  \hspace{1.1cm}\epsfig{file=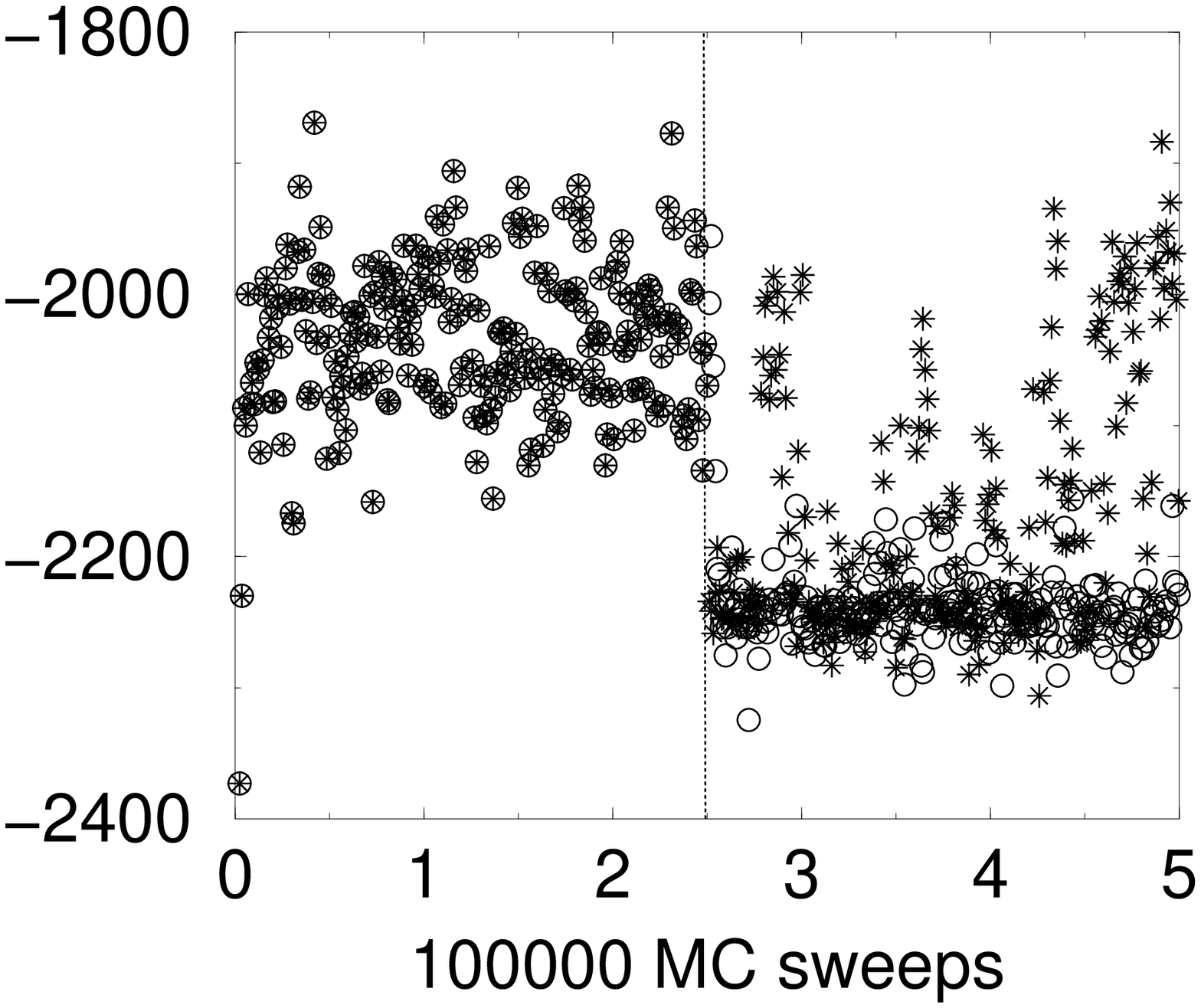,width=8.5cm}\\

  \vspace{1cm}
  \epsfig{file=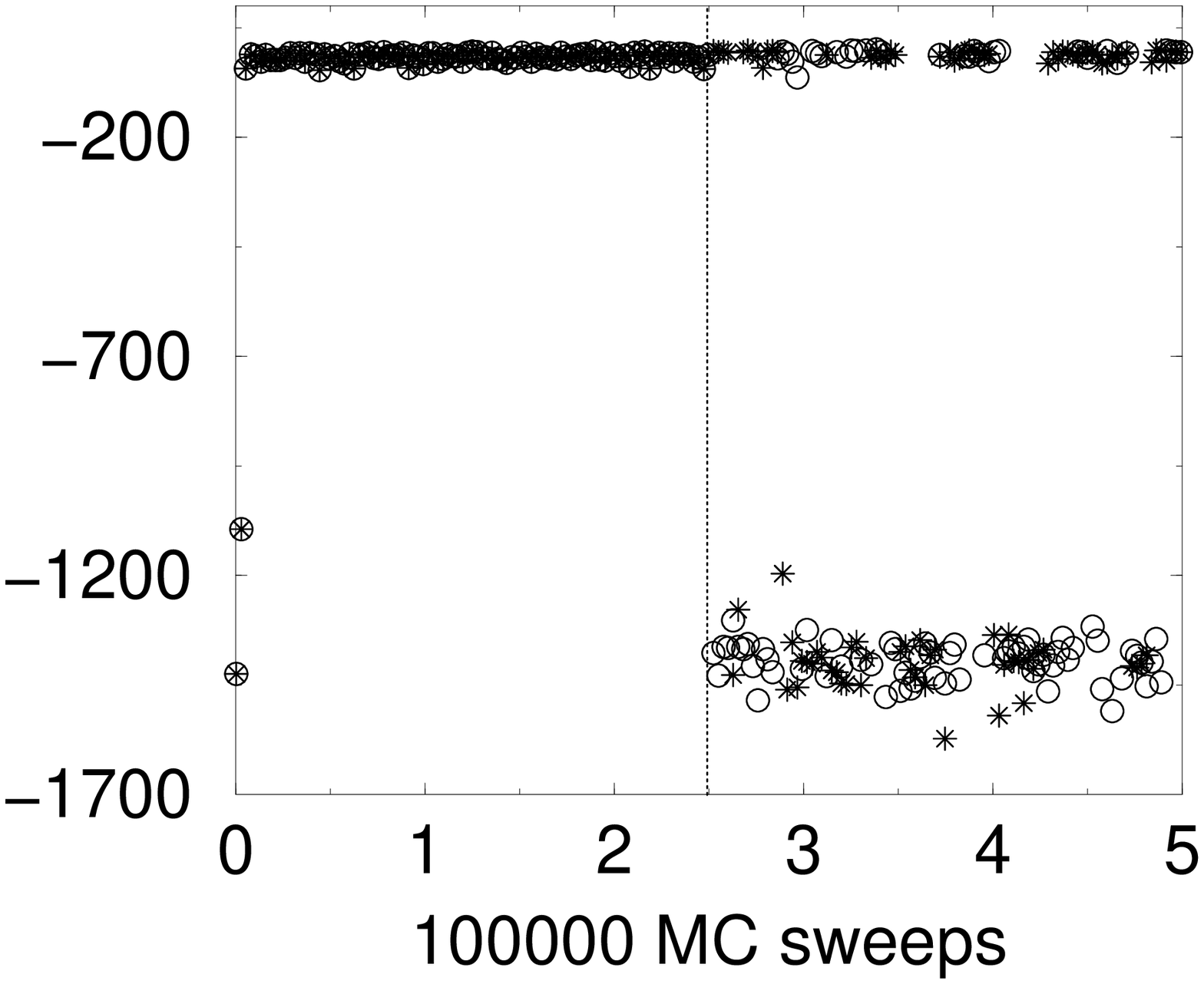,width=10cm}
  \caption{Upper: Energy of the lowest temperature replica (T=0.79) as a 
function of Monte Carlo sweeps. Circles: inversion temperature T$^{invert}$=1,
Stars: T$^{invert}$=T$_{min}=0.79$. After 250,000 steps (doted line) the 
first umbrella is incorporated in the simulation. The high inversion 
temperature umbrella leads to a complete suppresion of sampling of the 
relevant energies. Lower: Same as the upper figure, but for the replica at
T=1.03. For clarity we only show data every 25 cycles.}
  \label{fig:invert}
\end{figure}

Different replicas are simulated at different temperatures; we must therefore 
choose the particular temperature at which to perform the operations involved 
in Eqn.~(\ref{eq:binv}). Note that if all the replicas were at the same 
temperature, the Boltzmann inversion would yield the free energy difference 
between iterations $n$ and $n+1$. In our case, however, it is a corrective
procedure for the weights employed in the simulation, which were designed to 
produce a flatter distribution. We find that inversion at the minimum 
temperature (which corresponds to the maximum $\beta$, denoted by 
$\beta_{\text{max}}$) provides an optimum choice. At first glance, inverting 
every histogram at the temperature at which it was generated and adding up the 
resulting umbrellas would appear to be the best choice. This implementation
was successful in the sense that it leads to flatter distributions, but it 
doubled the number of iterations required to achieve the same accuracy as that 
attained by inversion at the lowest temperature. This is due to the fact that 
the error of the histograms is highest in the flanks. Adding up two umbrellas 
with large errors but correct temperature is less efficient than adding
up the histograms of different temperatures and inverting the global histogram.
Tests conducted by inverting at intermediate 
temperatures always overestimated the low temperature barriers, leading to
insufficient sampling in formerly highly frequented areas. In some cases the 
histograms were even shifted to completely new areas; Figure~\ref{fig:invert} 
illustrates that effect. In that case, a low-temperature replica (T=0.79) was 
shifted into a region of little interest by inverting at a temperature in the 
middle of the range covered by all the replicas (T=1). The region of little
interest corresponds in this case to energies which are only relevant at much
lower temperatures; the energies that we set out to sample were no longer
visited. Alternatively, an even lower temperature (lower than
the minimum temperature of the simulation) could be used for inversion; in that
case, however, the efficiency of the algorithm deteriorates considerably, as 
not even the lowest-temperature histogram becomes sufficiently flat.

In the first iteration, the Boltzmann weight $-\beta_i E+\beta_i\mu_i N$ is 
exchanged by $-\beta_i(E+\xi^{(1)}(E))+\beta_i\mu_i N$.The change of weights 
between two successive iterations $n$ and $n+1$ can therefore be expressed as
\begin{eqnarray}
  w_{\text{mc}}^{(n+1)}(E) &=& \frac{w_{\text{mc}}^{(n)}(E)}
 {\Omega^{\prime}(E)} = \nonumber
  w_{\text{mc}}^{(n)}(E)\exp\{-\beta [\xi^{(n+1)}(E)-\xi^{(n)}(E)]\}\\
  &=&w^{(0)}(E)\exp\{-\beta \xi^{(n+1)}(E)\},
\end{eqnarray}
where $w^{(0)}$ denotes the original, grand-canonical ensemble weight (i.e. 
$w^{(0)}=\exp(-\beta E +\beta\mu N)$). For the swap moves in the Metropolis 
criteria of Equation~(\ref{eq:metropolis}), the energy $E$ is interchanged
with the umbrella-corrected energy, $E+\xi(E)$.
\section{Results for the Lennard-Jones fluid}
The Lennard-Jones fluid provides a standard test for the study of new methods,
partly because highly accurate data are available for
comparison~\cite{smit92,wilding95,yan99}. In this work, we apply the
multicanonical parallel tempering method described above to determine its
vapor-liquid phase behavior. The interaction potential is given by
\begin{eqnarray}
 \Gamma(r)&=&4\epsilon\left[\left(\frac{\sigma}{r}\right)^{12}-
    \left(\frac{\sigma}{r}\right)^6\right]\,,\,r<r_c=2.5\sigma \\
  \Gamma(r) &=& 0 \,,\,\,\ r\ge r_c.
\end{eqnarray}
All data are reported in standard dimensionless units (distances are measured 
in $\sigma$, energies and temperature in $\epsilon$). The box length is 
$L=8\sigma$; the grand-canonical ensemble is used as the parent ensemble for 
the simulation. The multicanonical umbrellas are one-dimensional and are only 
functions of the energy $\xi=\xi(E)$. A two-dimensional implementation of the 
umbrella $(\xi=\xi(E,N))$ was also attempted, but this deteriorated the overall
performance of the algorithm and is not presented here. After several days of 
computer time, our two-dimensional implementation of the algorithm had not 
converged. When two-dimensional umbrellas are employed, the initial runs must 
be much longer to fill the histograms and reduce their statistical uncertainty.
Furthermore, addressing a two-dimensional table (as opposed to a simple vector)
hinders computational performance. A possible solution to this problem could 
perhaps be found through the two-dimensional generalization~\cite{yan01sb} of 
the recently proposed Density of States Monte Carlo method~\cite{wang01a}. That
method, however, is in its early stages of development and it is unclear 
whether it will work for large systems.

The preliminary runs to produce the umbrella were performed with between 
$50 * 10^3$ and $250 * 10^3$ MC cycles for every $\xi^{(i)}(E)$. A swap between
neighboring replicas was attempted every 100 Monte Carlo cycles. The last 60\%
of these cycles were used to create the energy histograms for inversion in the
following step. Equilibration was determined by the decay of the
auto-correlation function of the energy, and by the fact that the histograms 
did not change significantly over time. The histograms generated using only 
$20*10^3$, $50 * 10^3$, or $250 * 10^3$ Monte Carlo were essentially 
undistinguishable.  The correlation time was a few hundred MC cycles, which 
corresponds to a few successful swaps. These findings suggest that in the 
particular case of a Lennard-Jones fluid (at liquid-like densities), $20*10^3$
Monte Carlo cycles are sufficient to equilibrate the system in the new umbrella
for subsequent histogram collection.

\begin{figure}
  \begin{minipage}{0.49\linewidth}
    \epsfig{file=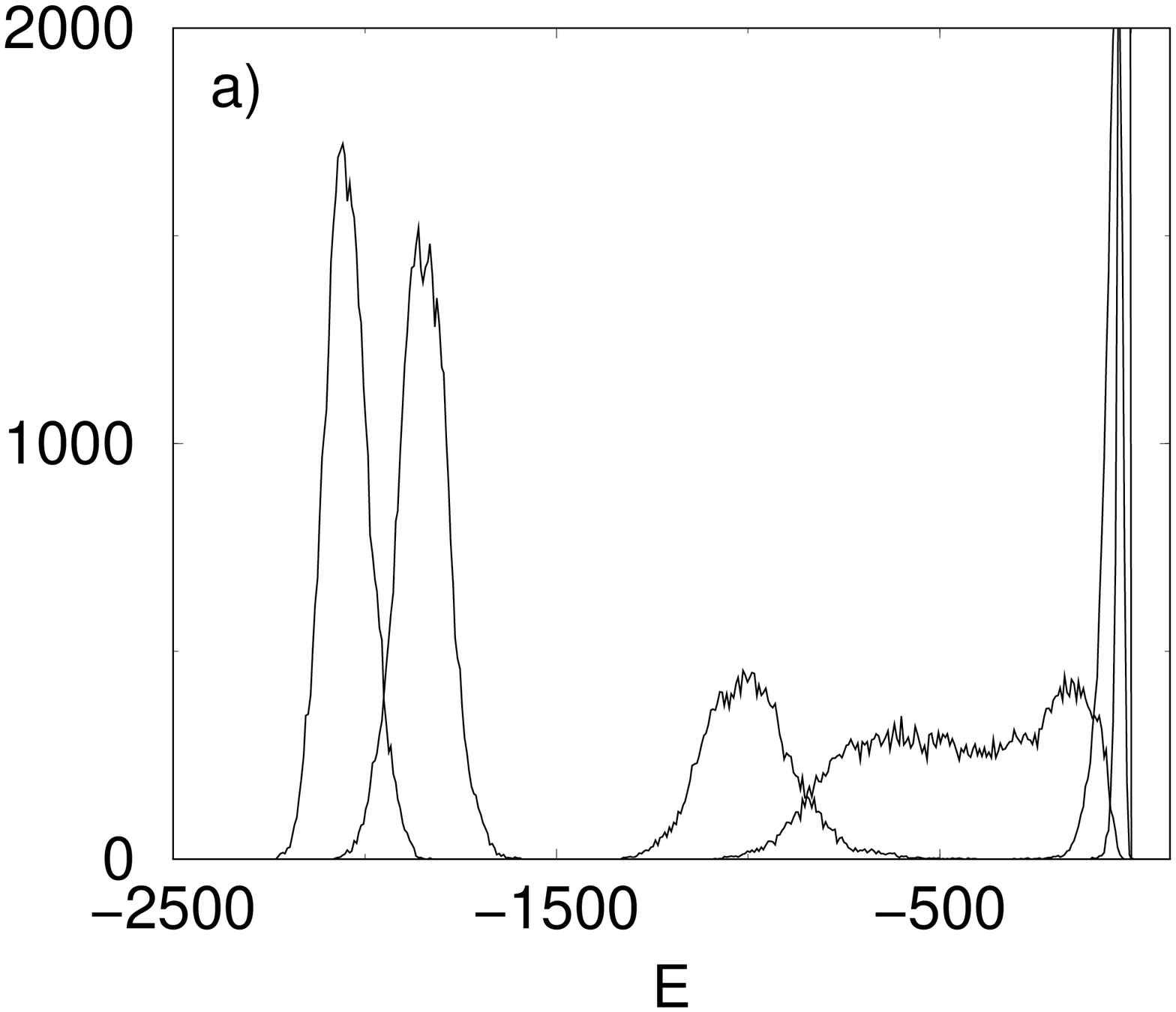, angle=-90,width=0.9\linewidth}
  \end{minipage}
  \begin{minipage}{0.49\linewidth}
    \epsfig{file=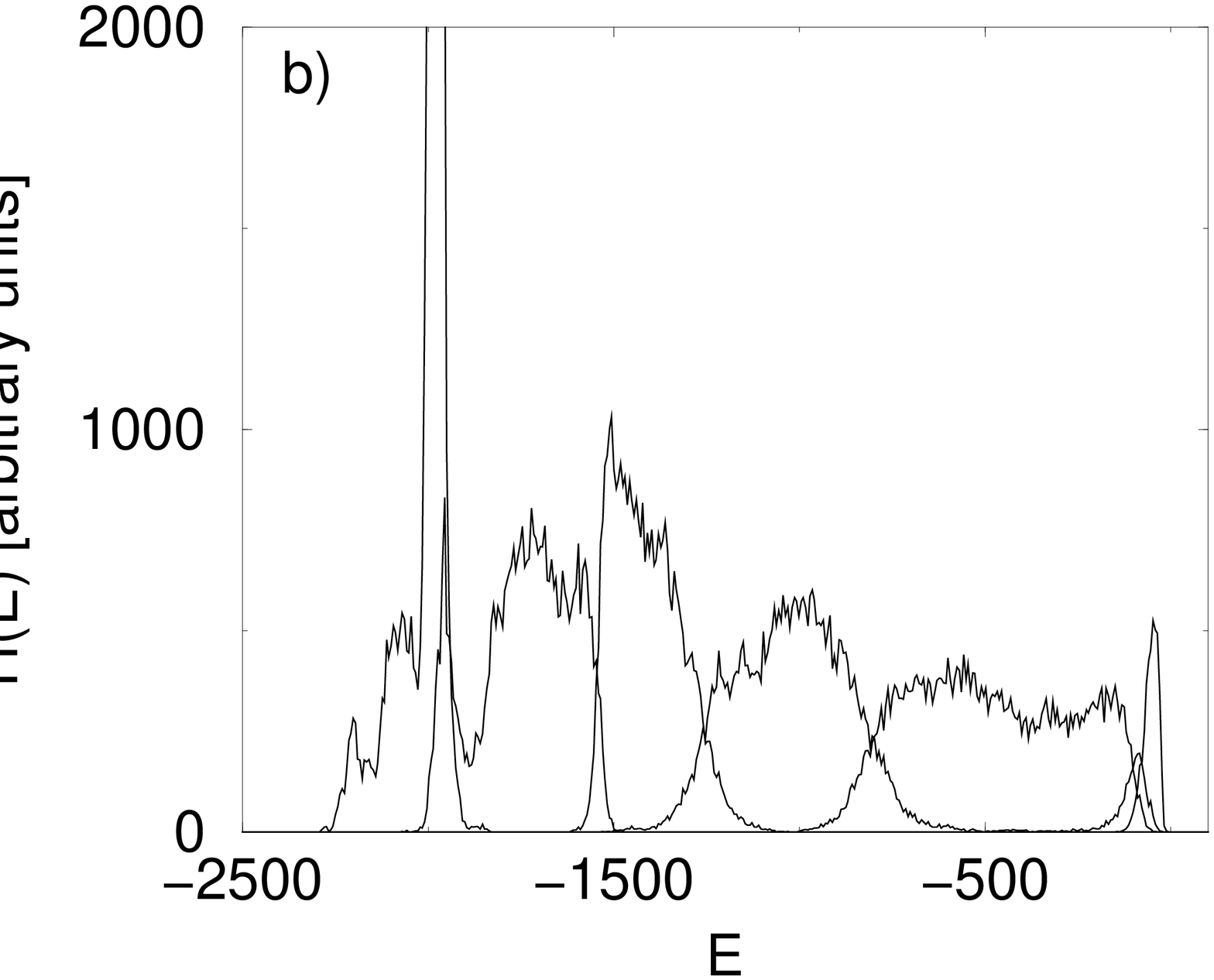, angle=-90,width=0.9\linewidth}
  \end{minipage}
  \caption{ Energy histograms with and without application of the 
  multicanonical weights used to calculate the Lennard-Jones phase diagram. 
  Part a) shows the histograms of the potential energy per replica without 
  multicanonical simulation, Part b) with application of the multicanonical 
  weights, which were produced in 4 iterations containing 250,000 MC cycles 
  each.}
  \label{fig:histo}
\end{figure}

The multicanonical weights are initially calculated as outlined above, until a
desired swap rate between neighboring replicas is achieved; a good acceptance 
probability is about 5 to 10\%. For six replicas, the sought-after overlap 
between neighboring histograms was attained after 4 iterations (see
Figure~\ref{fig:histo}). Note that, by design, the energy histograms 
corresponding to our original choice of state points 
(Table~\ref{tab:conditions}) did not overlap with each other. 
\begin{table}
  \begin{center}
    \begin{tabular}{rrr}
    \hline
    $i$ & T$_i$ & $\mu_i$\\
    \hline
    1 & 0.79 & $-4.76$ \\
    2 & 0.86 & $-4.23$ \\
    3 & 1.03 & $-3.35$ \\
    4 & 1.11 & $-3.02$ \\
    5 & 1.20 & $-2.74$ \\
    6 & 0.94 & $-3.98$ \\
    \hline
  \end{tabular}
  \caption{Thermodynamic conditions for a grand-canonical ensemble tempering
   simulation using 6 replicas. These conditions are taken as starting points
   for the multicanonical optimization. These thermodynamic conditions are
   taken from the literature~\cite{yan99}.}
\label{tab:conditions}
\end{center}
\end{table}
The temperatures
and chemical potentials were chosen as a subset from earlier parallel tempering
simulations with 18 replicas~\cite{yan99}. Conventional parallel-tempering swap
moves would not work for that choice. After applying the multicanonical 
weights, overlap between all replica histograms was achieved, thereby leading 
to a highly effective parallel tempering simulation. We performed two 
independent simulations of $10^6$ steps; only data produced over the last three
quarters of the simulation were used for analysis. Note that in cases where we
used a different inversion temperature to produce the umbrellas, or in cases 
were every histogram was inverted at its temperature (see above), we did not 
find any significant changes in the results; the production of the histograms,
however, was considerably slower.
As a word of caution, we point out that it is important that the umbrella 
potentials be based on well equilibrated histograms. This can cut down the 
number of iteration steps significantly, especially in the initial stages of 
the process where the weights change drastically between successive iterations.

In order to determine the optimal distribution of state points, we also 
performed simulations using 4 and 8 state points (or replicas) for 
multicanonical parallel tempering simulations. For reference, we also conducted
a conventional canonical parallel tempering simulation with 12 and 18 state 
points. Again, the state-points were subsets taken from ref.~\cite{yan99}. A 
one replica, purely multicanonical simulation was attempted for reference; that
simulation was not successful in covering the entire region of interest (with 
a single replica) and the initial runs did not converge. The number of 
iterations needed for the calculation of the umbrellas is shown in 
Table~\ref{tab:effi}, along with the time required for production runs. We find
that with 4 replicas we needed an unreasonably high number of iterations, and 
with 8 replicas we could not cut down on the number of iterations required with
respect to 6 replicas. For the size of our simulated system it appears that 6 
replicas provide an optimal choice. It is important to emphasize, however, 
that the optimum is likely to vary significantly depending on the particular 
details of a given system. Of course, the time required for the final 
production run scales with the number of replicas; from this point of view, 
the proposed algorithm is three times more efficient than the original 
18-replica multidimensional parallel tempering simulation of the same 
system~\cite{yan99}.
\begin{table}
  \begin{center}
    \begin{tabular}{cccc}
    \hline
    \mc{\# replicas} & \mc{\# iterations} &\mc{\# of iterations $\times$
    \# replicas} & {CPU time}\\
    N & $N_u$ & N$\cdot N_u$& $T_{\text{CPU}}$/h\\
    \hline
     1 & $> 80$&$>80$ & $>80$ \\
     4 & 11 & 44 &  76\\
     6 &  4 & 24 &  72\\
     8 &  5 & 40 &  80\\
    12 &  - & -  & 120\\
    18 &  - & -  & 180\\
    \hline
    \end{tabular}
    \caption{Number of iterations used for calculation of the umbrellas
    $N_u$, number of iterations times number of replicas (N$\cdot N_u$), and
    overall CPU time $T_{\text{CPU}}$ (in hours) required to simulate the
    relevant system on a simple personal computer
    (800 MHz AMD processor). For the case of one replica, we were unable
    to produce an individual umbrella capable of covering the entire
    coexistence region.}
    \label{tab:effi}
  \end{center}
\end{table}

The new method permits coverage of the entire coexistence region using only 6 
simulation replicas. One additional, important advantage over a traditional parallel 
tempering simulation resides in the smaller size of the overall system, which 
reduces drastically memory requirements. For systems like glasses or polymers, 
a large number of replicas would be necessary for parallel tempering 
simulations; such calculations would require extraordinary computational
resources, as the amount of memory scales linearly with the number of replicas.
For a large polymeric system, an individual replica containing several tens of
thousand interaction sites is already at the limits of a standard personal 
computer or workstation. With the combined parallel-tempering-multicanonical 
approach proposed here, such systems can be simulated using a smaller number of
replicas, thereby reducing the computational requirements considerably (both 
memory and production run time.)

\begin{figure}
 \epsfig{file=phasefit.epsi, angle=-90,width=8.5cm}\\
  \epsfig{file=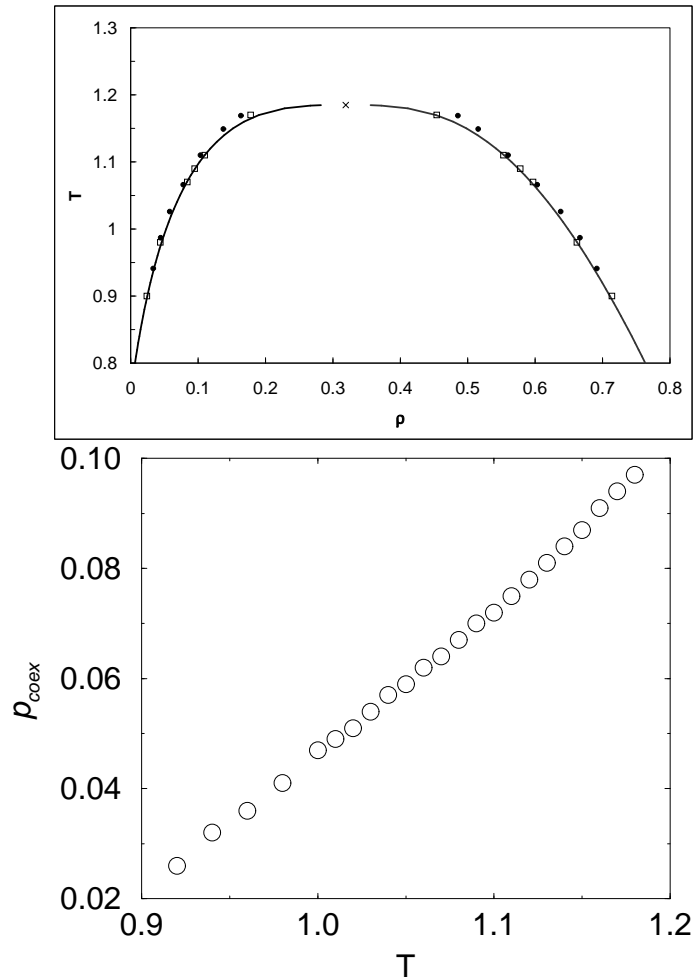,  angle=-90,width=8.5cm}
  \caption{Upper: Calculated Phase Diagram for the Lennard-Jones fluid. The
   open squares show results of this work. The circles show literature
   data~\cite{wilding95}. The solid line is an Ising fit to the data. Lower:
   Coexistence (vapor) pressure as a function of temperature for the same 
   system.}
\label{fig:phasediag}
\end{figure}

Figure~\ref{fig:phasediag} shows the phase diagram and the coexistence pressure
as a function of temperature calculated using the above-mentioned six replicas.
We use multi-histogram reweighting to calculate the other 
state-points~\cite{ferrenberg88,ferrenberg89}. As there is good overlap in the
histograms the accuracy of the reweighting procedure increases. The
phase-diagram is in good agreement with literature data~\cite{wilding95,yan99}.
The estimated critical temperature and density are
$T_c=1.18\pm0.01$, $\rho_c=0.32\pm0.02$. These were obtained by an
Ising fit to the reweighted data. We did not apply a finite-size
scaling correction.
\section{Conclusions}
We have presented an implementation of multicanonical sampling in the framework
of parallel tempering simulations. This combination of methods decreases the
number of necessary replicas and increases the efficiency of the simulation.
The validity of the new method has been demonstrated in the context of a simple
Lennard-Jones fluid, for which high-accuracy literature data for the 
coexistence curve are reproduced.

As pointed out in the multicanonical-ensemble literature, we note that it is 
important to exercise care in the selection of multicanonical weights. However,
when proper precautions are followed, the result is an effective methodology. 
Even for simple systems, such as the Lennard-Jones fluid, the memory and time
required to produce results over a wide range of conditions can be decreased by
a factor of three without an overhead to the overall calculation. We expect the
proposed formalism to be particularly well suited for simulation of systems 
which, by their very nature, require large sizes (e.g. polymers). An 
additional, important advantage of the proposed method is that it provides a 
remarkably simple way of parallelizing an umbrella-sampling simulation; to the
best of our knowledge, the implementation presented here constitutes the first
version of an umbrella sampling simulation that is parallelized.
\section*{Acknowledgments}
RF thanks Markus Deserno for fruitful discussions. Financial support from the
Petroleum Research Fund administered by the American Chemical Society, the
National Science Foundation (CTS-9901430), and the Emmy-Noether Program of the
German Research Foundation (DFG) is gratefully acknowledged.
\bibliography{standard}
\bibliographystyle{aip}

\end{document}